\documentclass[longnamesfirst,apjl]{emulateapj}
\usepackage{apjfonts,natbib}
\newcommand{\beq}{\begin{equation}}
\newcommand{\eeq}{\end{equation}}
\newcommand{\bea}{\begin{eqnarray}}
\newcommand{\eea}{\end{eqnarray}}
\begin{document}
\title{On the Origin of Nuclear Star Clusters in Late Type Spiral Galaxies}
\author{Milo\v s Milosavljevi\'c $^1$}
\affil{$^1$Theoretical Astrophysics, California Institute of Technology, Pasadena, CA 91125, milos@tapir.caltech.edu}
\shorttitle{NUCLEAR STAR CLUSTERS}
\shortauthors{MILOSAVLJEVI\'C}

\begin{abstract}
A large fraction of bulgeless disk galaxies contain young compact stellar
systems at their centers, in spite of the local gravitational stability of
these disks.  We evaluate two contrasting hypotheses for the origin of the
nuclear star clusters in late-type disk galaxies.  The
clusters could not have migrated from distant eccentric locations in the
disk.  Instead they must have formed in situ, requiring radial transport
of gas toward the center of the disk.  This transport could be a
consequence of the development of the magnetorotational instability in the
differentially rotating warm neutral medium.  We evaluate the
rate of gas transport into the disk center and find that 
it is sufficient to support continuous star formation in that location.  
Enhanced stellar surface brightness in the inner few hundred parsecs and
the formation of a compact stellar system in the central few parsecs are
unavoidable in dark matter halos with divergent density profiles.  We
illustrate our conclusions on a model of the nearest late-type disk galaxy
M33.
\end{abstract}


\section{Introduction}
Recently, interest has been growing in the compact stellar clusters found in the centers of late-type spiral galaxies \citep{Carollo:97,Boker:99,Matthews:99,Boker:01,Boker:02,Matthews:02,Boker:03a,Schinnerer:03,Boker:03b}.  Imaging with the {\it Hubble Space Telescope} of the central regions of $\sim75\%$ \citep{Boker:02} of late-type spirals (Hubble type Scd or later) has revealed young, compact stellar systems with luminosities $-14\lesssim M_I\lesssim-9$ that dominate the photocenters of these disk galaxies possessing no discernable stellar bulge. The ``nuclear star clusters'' (NSCs) have half-light radii $R_{\rm cl}\lesssim 5\textrm{ pc}$ and luminosity masses $\sim10^{4}-10^{7}M_\odot$ \citep{Boker:02,Boker:03b}.  The dynamical masses of the NSCs, $\sim 10^{6}-10^{8}M_\odot$, are an order of magnitude higher than would be expected from stellar synthesis models, indicating the presence of older stellar generations that have faded \citep{Walcher:04}\footnote{See http://www.ociw.edu/ociw/symposia/series/symposium1/proceedings.html.}. Some of the NSC host galaxies exhibit spiral structure, dust lanes, and other morphological complexities, but many do not, leading \citet{Boker:02} to suggest that the NSC phenomenon is not related to nonaxisymmetric features in the galactic disk.  \citet{Boker:03b} find no evidence for a correlation between the presence of a cluster and the presence of a large-scale stellar bar. Stellar disks of the host galaxies have exponential surface brightness profiles to within hundreds of parsecs of the center.  Surface brightness profiles of the inner few hundred parsecs but outside the NSC frequently exhibit brightness excess above the exponential law \citep{Boker:03a}.

The formation mechanism of NSCs is unknown. The central rotation curves of some Scd galaxies rise linearly with radius \citep{Matthews:02}, implying near solid-body rotation and a shallow potential well.  Rotation curves of NSC host galaxies resemble those of low surface brightness and dwarf spiral galaxies.  Radial mass profiles of the latter two classes of galaxies have been extensively studied (e.g.,~\citealt{deBlok:03,Swaters:03,Simon:03}). Models of the combined dark matter and baryonic density profiles that are finite at the center (e.g.,~\citealt{Burkert:95}), as well as those that are mildly divergent $\rho\propto r^{-\gamma}$ with $\gamma\lesssim 1$ (e.g.,~\citealt{Navarro:97}), both yield admissible fits to the rotation curves, while steeper profiles are ruled out. 

Two scenarios for the origin of NSCs are possible. They could have assembled in situ from the material available in the central region of the disk. Alternatively, they could have formed remotely and have subsequently migrated to the center via dynamical friction.   We test these scenarios on a model (\S~\ref{sec:m33}) of the closest and the well-studied NSC host galaxy M33.  In \S~\ref{sec:insitu}, we argue that the first scenario---in situ formation---is the likely origin of NSCs.  In \S~\ref{sec:infall} we demonstrate that the alternative migratory scenario fails to deliver NSCs to their observed locations, and we discuss observable implications of our model.

\section{A Model Host Galaxy: M33}
\label{sec:m33}

The nucleus of M33 contains a compact star cluster that has elicited attention for a decade (e.g., \citealt{Kormendy:93b,Lauer:98,Matthews:99,Davidge:00}) because of its resemblance to the massive black hole inhabited nucleus of the Galaxy.  No massive black hole, however, has been found in M33 \citep{Merritt:01,Gebhardt:01}.  The interstellar medium and the magnetic field of M33 have also been extensively studied (e.g.,~\citealt{Wilson:89,Engargiola:03,Beck:00}).  Rotation curves of the more distant NSC host galaxies appear to be rising even more slowly \citep{Matthews:02} than that of M33 \citep{Corbelli:00}.

We adopt the following parameters for M33: present NSC mass $M_{\rm cl}\sim10^5M_\odot$ \citep{Lauer:98,Long:02}, surface density of \ion{H}{1} of $\Sigma\sim10M_\odot\textrm{ pc}^{-2}$ \citep{Engargiola:03}, \ion{H}{1} disk scale length $R_0\sim10\textrm{ kpc}$ \citep{Engargiola:03}, average total magnetic field strength $B\sim6\ \mu\textrm{G}$ \citep{Beck:00}, and a thermal sound speed of $c_s\sim10\textrm{ km s}^{-1}$.

\section{Formation of Nuclear Star Clusters In Situ}
\label{sec:insitu}

Given a dynamical mechanism that transports gas from larger to smaller radii in the disk, an NSC can form from the gas that accumulates at the center of the disk.  The source of radial transport in non self-gravitating disks is the magnetic stress if the magnetorotational instability (MRI) develops in the disk.  When there is differential rotation, torques arise between concentric rings of interstellar gas.  The Alfv\'en velocity in a fluid with density $\rho$, velocity ${\bf u}$ relative to the circular orbit motion, and magnetic field ${\bf B}$ is given by ${\bf u}_A=(4\pi\rho)^{-1/2}{\bf B}$.  The few microgauss magnetic field measured in the disks of galaxies such as M33 is compatible with being in equipartition with the observed velocity dispersion ($\sim10\textrm{ km s}^{-1}$), where the density is estimated via $\rho\sim\Sigma v/c_s R$ ($v$ is the velocity of the rotation curve of the galaxy at radius $R$). \citet{Sellwood:99} pointed out that the velocity dispersion could be a result of magnetohydrodynamic turbulence. 

Ignoring self-gravity of the fluid, the stress tensor  $T_{R\phi}=\langle\rho(u_R u_\phi-u_{AR} u_{A\phi})\rangle$ is an average of Reynolds and Maxwell stresses, where we have adopted cylindrical coordinates $(R,\phi,z)$. Correlations between velocity and magnetic field components, which give rise to Reynolds and Maxwell stresses, induce an outward angular momentum flux and an inward mass flux \citep{Balbus:98}.  The mass-weighted vertically averaged specific stress tensor is given by $W_{R\phi}=T_{R\phi}/\langle\rho\rangle$ and has dimensions of square velocity.
 
We parameterize the average stress tensor as $\langle W_{R\phi}\rangle=\alpha c_s^2$, where $\alpha$ is a parameter introduced by \citet{Shakura:73} that can be determined by measuring the stress-to-pressure ratio in numerical simulations.  We assume that $\alpha$ and $c_s$ are weak functions of radius.  The accretion rate is then given by (e.g.,~\citealt{Frank:02})  
\beq
\dot M=\frac{4\pi \alpha c_s^2 \Omega}{\kappa^2 R} \frac{\partial}{\partial R} (R^2\Sigma) ,
\eeq 
where $\kappa=[R^{-3}d/dR(R^4\Omega^2)]^{1/2}$ is the epicycle frequency. 

The values of $\alpha$ associated with the MRI have been measured in three-dimensional simulations (e.g.,~\citealt{Hawley:95,Stone:96,Armitage:98}).  For the relevant isothermal equation of state, $\alpha\sim0.1$ when the initial (weak) magnetic field is uniform and perpendicular to the disk plane and $\alpha\sim0.01$ when it is nonuniform.  Recently, \citet{Kim:03} conducted shearing-box simulations of a section of a gaseous galactic disk with an initial weak vertical magnetic field and measured $\alpha\approx0.15-0.3$. 

Atomic gas in a galactic disk can be treated as an ideal magnetic fluid if a sufficient fraction of ions are interspersed among the neutral atoms to provide for an inertial coupling of the neutrals to the magnetic field.  \citet{Blaes:94} showed that adequate coupling is achieved when the collision frequency of a neutral with the ions exceeds the local epicyclic frequency, $\tilde\gamma\rho_i\gtrsim\kappa$, where $\tilde\gamma=(m_n+m_i)^{-1}\langle\sigma v\rangle_{ni}+\zeta/\rho_i$ is expressed in terms of the density of ions $\rho_i$, the momentum rate coefficient $\langle\sigma v\rangle_{ni}\approx 1.9\times10^{-9}\textrm{ cm}^{3}\textrm{ s}^{-1}$ \citep{Draine:83}. Here, $m_n$ and $m_i$ are, respectively, the masses of the neutrals and the ions (taken here to be the hydrogen mass), and $\zeta$ is the total ionization rate per molecule.  

\citet{Wolfire:03} argue that in the Galactic disk, the ionizing flux is made of extreme UV photons, soft X-rays, and cosmic rays, with a total of $\zeta\sim 5\times10^{-17}\textrm{ s}^{-1}$.  They also evaluate the ionization fraction as a function of the density of neutrals, $\chi_e(n)\sim 10^{-3} (n/10\textrm{ cm}^{-3})^{-1}$, consistent with earlier estimates (\citealt{Kulkarni:87} and references therein).  In \ion{H}{1}, the first term in the expression for $\tilde\gamma$ dominates.  This implies that the MRI can develop if $\kappa\lesssim2\times10^{-11}\textrm{ s}^{-1}$.  Using the available rotation curve data for M33 and borrowing the ionization fraction from the Galaxy, we estimate that $\kappa\sim10^{-12}\textrm{ s}^{-1}$ at $R=1\textrm{ pc}$ and $\kappa\sim3\times10^{-15}\textrm{ s}^{-1}$ at $R=1\textrm{ kpc}$.  (Solid-body rotation would correspond to the maximum epicycle frequency of $\kappa\sim3\times10^{-12}\textrm{ s}^{-1}$.)  Therefore $\tilde\gamma\rho_i\ll\kappa$ throughout the disk, implying that ideal magnetic fluid is a good approximation to \ion{H}{1} in disks of Scd galaxies, provided that conditions in these disks are similar to those in the Galactic disk.

Our goal is to evaluate the gas mass accumulated at the center of the disk in time $\Delta t$.  We assume that the total dynamical mass enclosed within radius $R$ in the center of the galaxy is a power-law of radius $M_{\rm dyn}(R)=R_1 V_1^2G^{-1}(R/R_1)^{3-\gamma}$, where again $\gamma=-d\log\rho/d\log r$. We have parameterized the dynamical mass---including dark matter, stars, and gas---by the circular velocity $V_1$ at some fixed radius $R_1$.  Circular velocity at a generic radius is $V(R)=V_1(R/R_1)^{1-\gamma/2}$ and thus $\Omega,\kappa\propto R^{-\gamma/2}$.  Surface density in the disk is in a quasi-steady state if the flux of mass through radius $R$ is independent of radius, $\partial\dot M_{\rm gas}/\partial R=0$.  This requires that $\Sigma\propto R^{-\gamma/2}$, and thus $\dot M_{\rm gas}=2\pi\alpha c_s^2 R_1 \Sigma(R_1) V_1^{-1}$.

The quasi-steady state is reached only at radii $R<R_{\rm acc}$ with accretion times $t_{\rm acc}\sim\pi R^2 \Sigma/\dot M_{\rm gas}$ shorter than the age of the disk $t_{\rm disk}$, where $R_{\rm acc}$ is the maximum such radius.  If we assume that the age of the disk of M33 is $t_{\rm disk}\sim 1\textrm{ Gyr}$ and that the viscosity parameter is $\alpha\sim 0.1$, we obtain $R_{\rm acc}\sim500\textrm{ pc}$.  

We compare the gas density profile $\Sigma_{\rm central}\propto R^{-\gamma/2}$ just inside $R_{\rm acc}$ to the profile $\Sigma_{\rm exp}\propto e^{-R/R_0}$ obtained by extrapolating the exponential density profile that describes the disk at larger radii unaffected by accretion.  Since $(\partial\Sigma_{\rm central}/\partial R)/(\partial\Sigma_{\rm exp}/\partial R)=\gamma R_0/2R_{\rm acc}$ at $R_{\rm acc}$, the density at $R<R_{\rm acc}$ exceeds the extrapolated density when $\gamma>2R_{\rm acc}/R_0\sim 0.1$.  If the star formation rate (SFR) per unit gas mass is uniform in the disk, the excess gas density in the central few hundred parsecs implies an excess in the surface brightness profile of the stellar light.

We proceed to evaluate the accreted mass.  The gas mass accumulating at the center of the disk in time $\Delta t$ equals
\bea
M_{\rm gas}&\sim&10^7 M_\odot 
\left(\frac{\Delta t}{10^9\textrm{ yr}}\right)
\left(\frac{c_s}{10\textrm{ km s}^{-1}}\right)^2
\left[\frac{\Sigma(R_{\rm acc})}{10M_\odot\textrm{ pc}^{-2}}\right]
\nonumber\\
& &\times
\left(\frac{\alpha}{0.1}\right)
\left(\frac{R_{\rm acc}}{500\textrm{ pc}}\right)
\left[\frac{V(R_{\rm acc})}{40\textrm{ km s}^{-1}}\right]^{-1}
 .
\eea
If $\sim10\%$ of this gas is converted into stars, a cluster forms with a total initial stellar mass of $\sim10^6M_\odot$.  This is compatible with the mass of a typical NSC.

Given that the development of MRI is contingent on differential rotation in the disk, do disks in near solid-body rotation exhibit the same central accumulation?  The instability can develop if the Alfv\'en speed is not larger than the vertical sound speed times a function quantifying the degree of differential rotation (see \citealt{Balbus:98}, eq. [110]). The condition for instability can be written as $u_A \lesssim (2/\pi) q^{1/2} c_s$, where $q=|d\log\Omega/d\log R|$.  

A disk could thus evade MRI 
if the rotation were solid body, i.e., if the total mass density were uniform. The growth time of the instability $2 \Omega^{-1} q^{-1}$, which here equals
$t_{\rm grow}\sim 5\times 10^7\gamma^{-1}(R/500\textrm{ pc})
(V/40\textrm{ km s}^{-1})^{-1}\textrm{ yr}$,
is short $t_{\rm grow}\ll t_{\rm disk}$ even in very shallow ($\gamma\gtrsim 0.1$) density profiles (note that $\rho\propto R^{-\gamma}$).  Therefore the development of MRI and its consequences---the radially inward transport and the central accumulation of gas---are universal in disks residing in singular host dark matter halo profiles, however mild that singularity may be.  After instability develops, the transport rate is independent of the dark halo density profile.

\section{NSC Could Not Have Migrated to the Center}
\label{sec:infall}
Here we evaluate the alternative scenario in which NSCs migrate to disk centers after forming elsewhere. We estimate the time for migration from radius $R$. Assuming that the host dark matter halo is spherical and concentric with the galactic disk, the total mass density inside the scale radius of the halo is 
$\rho({\bf r})=\rho_0 (r/r_0)^{-\gamma} + \Sigma_0 \exp(-R/R_0) \delta(z)$,
where $R=(x^2+y^2)^{1/2}$ and $\gamma$, $\rho_0$, $r_0$, $\Sigma_0$, and $R_0$ are constant parameters.  Assume that the cluster of mass $M_{\rm cl}$ forms on a disk-planar circular orbit at distance $R$ from the center.  The timescale for orbital decay via dynamical friction is given by $t\sim v_{\rm circ}^3/4\pi G^2 M \rho_<\ln\Lambda$ \citep{Chandrasekhar:43}, where $v_{\rm circ}$ is the circular velocity, $\rho_<$ denotes the density in dark matter particles moving slower than $v_{\rm circ}$, and $\ln\Lambda$ is the Coulomb logarithm.  The ratio $\rho_</\rho$ is unknown as it depends on the phase-space structure of the dark halo.  To derive a lower bound on the decay time we optimistically set $\rho_<=\rho= (3-\gamma)v_{\rm circ}^2/4\pi Gr^2$.  Furthermore, \citet{Velazquez:99} found that $\ln\Lambda\sim2$ for satellites sinking in galaxy disks with $\gamma=1$ halos. The decay time scale is then given by
\beq
t\gtrsim 3\times10^9\textrm{ yr}
\left(\frac{v_{\rm circ}}{50\textrm{ km s}^{-1}}\right)
\left(\frac{r}{1\textrm{ kpc}}\right)^2
\left(\frac{M_{\rm cl}}{10^6M_\odot}\right)^{-1} .
\eeq

Dynamical friction due to the disk is a more delicate issue.  Because of the linear response of disks to perturbations (e.g.,~\citealt{Goldreich:79}), positive and negative torques simultaneously act on the satellite.  Negative torque is exerted by the horseshoe orbits in the disk and amounts to \citep{Quinn:86}
\beq
\label{eq:quinn}
\Gamma\approx 2|AB| H^4 R\ \ d\Sigma/dR ,
\eeq
where $A=-d\Omega/d\log R$ and $B=A-\Omega$ are the Oort parameters, and $H=(GM_{\rm cl}/4A^2)^{1/3}$ is the radius of the Hill sphere centered on the cluster within which the cluster potential dominates differential flow in the disk.  

Equation (\ref{eq:quinn}) is valid when the Hill radius is larger than the disk scale height $h$. For a  smaller $M_{\rm cl}$ with $H<h$, the torque is smaller by fraction $(H/h)^2$. If the rotation curve is ``flat'' $\Omega\propto R^{-1}$, this result coincides with the detailed contribution from corotation resonances evaluated by \citet{Donner:93}.  Higher order resonances also contribute negative torque terms when $R\lesssim R_0$ but their contribution is comparable or smaller \citep{Wahde:96} than that of the corotation resonances and can be ignored.

The inspiral time due to the torque in equation (\ref{eq:quinn}) is
\bea
\label{eq:timequinn}
t&\approx& 2\times10^{10}\textrm{ yr} \frac{(4-\gamma) \gamma^{5/3}}{2-\gamma}
\left(\frac{v_{\rm circ}}{50\textrm{ km s}^{-1}}\right)^{5/3}
\left(\frac{R}{1\textrm{ kpc}}\right)^{-2/3}\nonumber\\
&\times&\left(\frac{\Sigma_0}{10M_\odot\textrm{ pc}^{-2}}\right)^{-1}
\left(\frac{M_{\rm cl}}{10^6M_\odot}\right)^{-1/3}
\left(\frac{R_0}{3\textrm{ kpc}}\right)e^{R/R_0} ,
\eea
where $\gamma$ henceforth denotes the negative logarithmic slope of the combined dark matter and baryonic density profiles.  Deceptively, the timescale in equation (\ref{eq:timequinn}) formally becomes infinitesimal when the disk is in near solid rotation $\gamma\rightarrow0$.  The implicit assumption $H\ll R$ breaks down in this limit, and we may try to replace $H$ with $R$ in equation (\ref{eq:quinn}) to obtain $t\sim t_{\rm dyn} M_{\rm cl}/M_{\rm gas}(R)$, where $t_{\rm dyn}=R/v_{\rm circ}$ is the dynamical time and $M_{\rm gas}$ is the enclosed gas mass.  However, at radii where $M_{\rm cl}/M_{\rm gas}(R)\sim 1$, the outward horseshoe orbits that provide the negative torque are depleted, and the net torque thus changes sign, thereby reversing the sense of the migration.

We contrast the exceptionally long timescales derived above with the estimated stellar ages of NSCs.  Using the bolometric luminosities and the density of stars on the asymptotic giant branch in the nucleus of M33, \citet{Stephens:02} inferred twin starbursts at $\sim0.5$ and $2$ Gyr.  These ages could be reconciled with dynamical friction facilitated infall only by starting with very small initial radii ($\ll 1\textrm{ kpc}$).  Furthermore, the best-fitting starburst models of \citet{Gordon:99} and \citet{Long:02} imply the presence of another, very recent ($\sim50-70\textrm{ Myr}$) star formation episode yielding a stellar mass of $\sim10^4 M_\odot$ or about $12\%$ of the total mass.  Similarly, \citet{Boker:99} determined the mean age of $10^{6.8}-10^{7.8}\textrm{ yr}$ for the nucleus of the giant Scd spiral IC 342 hosting a $6\times10^6M_\odot$ cluster. \citet{Davidge:02} infer enhanced star formation in the recent $0.1\textrm{ Gyr}$ in the nuclei of nearby Sc galaxies NGC 247 and NGC 2403.  This evidence suggests that some NSCs are much younger (or contain a much younger component) than the migration times, and we thus reject the migratory hypothesis. 

\section{Discussion and Conclusions}

When present, nonaxisymmetric features (bars and spiral structure) channel gas toward the center of the galaxy \citep{Shlosman:89}.  However, some NSC host galaxies exhibit negligible departures from axisymmetry.  Self-gravity in a disk gives rise to an effective kinematic viscosity (e.g.,~\citealt{Lin:87b}).  Stability of disks against self-gravity requires $Q>1$ where  $Q=c_s\kappa/\pi G\Sigma$ \citep{Toomre:64}.  The disks of NSC host galaxies are largely stable. Although that of M33 contains approximately equal masses in stars and gas, \citet{Corbelli:00} infer $Q>1$ throughout, while our model implies $Q\gg1$.  Failure of the Toomre criterion to explain the SFR in M33 is well known \citep{Wilson:89,Martin:01}.  Furthermore, \citet{Engargiola:03} argue that giant molecular clouds (GMCs) in M33 are transients forming directly from the atomic gas with a characteristic mass $\sim7\times10^4M_\odot$, which is much smaller than the Jeans mass $c_s^4/G^2\Sigma\sim3\times10^7M_\odot$ of the disk.  Therefore the GMCs are not like projectiles that facilitate radial transport through cloud-cloud interactions (e.g.,~\citealt{Silk:81}).

We find that the radial transport of interstellar gas in galactic disks susceptible to MRI is sufficient to facilitate an accumulation of gas in the central few hundred parsecs and the formation of a compact stellar system at the disk center.  The sizes and the ages of the central stellar systems in this model correspond closely to those of the NSCs in Scd galaxies. 

We speculate that after a cluster forms in the disk, the supply of gas flowing toward the nascent cluster is temporarily interrupted by the stellar winds and the supernova blast waves.  During the life time of O and B stars $\sim5\times10^7\textrm{ yr}$, the stellar energy output fuels a hot low-density ``superbubble'' centered on the cluster. The radius of the bubble reaches about the disk scale height $\sim 100\textrm{ pc}$ (e.g.,~\citealt{MacLow:88}).  After the stars with masses $\gtrsim8M_\odot$ have been expended, fresh gas returns to the bubble and accumulates toward a subsequent starburst.  The star formation history of an NSC is therefore punctuated by periodic starbursts separated by $\gtrsim10^8\textrm{ yr}$.

The radial transport of neutral gas in the disk implies a surface density enhancement inside a radius $R_{\rm acc}$ measuring hundreds of parsecs.  Remarkably, a ``central light excess'' a few hundred parsecs in radius, attributable neither to the NSC nor to the exponential stellar disk, has been observed in several host galaxies (B\"oker et al.~2003) including M33 \citep{Kent:87}.  The excess is a natural consequence of the transport.  Indeed, while the excess is observed in the stellar light and it is the gas that is transported by the MRI, the SFR in the overdense region may well trace gas density as elsewhere in the disk ($Q$ is independent of radius for $R\leq R_{\rm acc}$; see \S~\ref{sec:insitu}).

Bright nuclear disk components of the late-type spirals devoid of luminous NSCs, such as in the superthin galaxy UGC 7321 \citep{Matthews:00}, could also be the result of radial transport of interstellar gas. 
We expect that 
the NSC luminosities depend strongly on the time elapsed
since the most recent starburst episode.  
As dynamical data on NSCs become available, 
it will be possible to estimate the period 
between, and the magnitude of the
consecutive episodes, which may lend 
insight into the assembly of bulges and ``pseudobulges'' \citep{Kormendy:93a}.

\acknowledgements
We thank T.~B\"oker, M.~Kamionkowski, and the anonymous referee for detailed comments on the manuscript and P.~Goldreich, J.~Hawley, L.~Ho, W.~Kim, A.~MacFadyen, S.~Phinney, and R.~Sari for valuable discussions.  This research was supported at Caltech by a postdoctoral fellowship from the Sherman Fairchild Foundation.


\begin{thebibliography}{99}

\expandafter\ifx\csname natexlab\endcsname\relax\def\natexlab#1{#1}\fi
\expandafter\ifx\csname url\endcsname\relax
  \def\url#1{{\tt #1}}\fi
\expandafter\ifx\csname urlprefix\endcsname\relax\def\urlprefix{URL }\fi
\providecommand{\eprint}[2][]{\url{#2}}

\bibitem[Armitage(1998)]{Armitage:98} Armitage, P.~J.\ 1998, \apjl,
501, L189 

\bibitem[\protect\astroncite{{Balbus} \& {Hawley}}{1998}]{Balbus:98}
{Balbus}, S.~A., \& {Hawley}, J.~F., 1998,
 { Rev. Mod. Phys.}, { 70}, 1

\bibitem[\protect\astroncite{{Beck}}{2000}]{Beck:00}
{Beck}, R., 2000,
 in Proc. 232 WE-Heraeus Seminar, The Interstellar Medium in M31 and M33, ed. E.~Berkhuijsen, R.~Beck, \& R.~Walterbos (Aachen: Shaker), 171

\bibitem[\protect\astroncite{{Blaes} \& {Balbus}}{1994}]{Blaes:94}
{Blaes}, O.~M., \& {Balbus}, S.~A., 1994,
 { \apj}, { 421}, 163

\bibitem[\protect\astroncite{{B\"oker} {\rm et~al.\/}}{2001}]{Boker:01}
{B{\" o}ker}, T., {van der Marel}, R.~P., {Mazzuca}, L., {Rix}, H., {Rudnick},
  G., {Ho}, L.~C., \& {Shields}, J.~C., 2001,
 { \aj}, { 121}, 1473

\bibitem[\protect\astroncite{{B\"oker} {\rm et~al.\/}}{2002}]{Boker:02}
{B{\" o}ker}, T., {Laine}, S., {van der Marel}, R.~P., {Sarzi}, M., {Rix}, H.,
  {Ho}, L.~C., \& {Shields}, J.~C., 2002,
 { \aj}, { 123}, 1389

\bibitem[B{\" o}ker et al.(2004)]{Boker:03b} B{\" o}ker, T.,
Sarzi, M., McLaughlin, D.~E., van der Marel, R.~P., Rix, H., Ho, L.~C., \&
Shields, J.~C., 2004, \aj, 127, 105 

\bibitem[\protect\astroncite{{B\"oker}, {Stanek} \& {van der Marel}}{2003}]{Boker:03a}
{B{\" o}ker}, T., {Stanek}, R., \& {van der Marel}, R.~P., 2003,
 { \aj}, { 125}, 1073

\bibitem[\protect\astroncite{{B\"oker}, {van der Marel} \& {Vacca}}{1999}]{Boker:99}
{B{\" o}ker}, T., {van der Marel}, R.~P., \& {Vacca}, W.~D., 1999,
 { \aj}, { 118}, 831

\bibitem[\protect\astroncite{{Burkert}}{1995}]{Burkert:95}
{Burkert}, A., 1995,
 { \apjl}, { 447}, L25

\bibitem[\protect\astroncite{{Carollo} {\rm et~al.\/}}{1997}]{Carollo:97}
{Carollo}, C.~M., {Stiavelli}, M., {de Zeeuw}, P.~T., \& {Mack}, J., 1997,
 { \aj}, { 114}, 2366

\bibitem[\protect\astroncite{{Chandrasekhar}}{1943}]{Chandrasekhar:43}
{Chandrasekhar}, S., 1943,
 { \apj}, { 97}, 255

\bibitem[\protect\astroncite{{Corbelli} \& {Salucci}}{2000}]{Corbelli:00}
{Corbelli}, E., \& {Salucci}, P., 2000,
 { \mnras}, { 311}, 441

\bibitem[\protect\astroncite{{Davidge}}{2000}]{Davidge:00}
{Davidge}, T.~J., 2000,
 { \aj}, { 119}, 748

\bibitem[\protect\astroncite{{Davidge} \& {Courteau}}{2002}]{Davidge:02}
{Davidge}, T.~J., \& {Courteau}, S., 2002,
 { \aj}, { 123}, 1438

\bibitem[\protect\astroncite{{de Blok} {\rm et~al.\/}}{2003}]{deBlok:03}
{de Blok}, W.~J.~G., {Bosma}, A., \& {McGaugh}, S., 2003,
 { \mnras}, { 340}, 657

\bibitem[\protect\astroncite{{Donner} \& {Sundelius}}{1993}]{Donner:93}
{Donner}, K.~J., \& {Sundelius}, B., 1993,
 { \mnras}, { 265}, 88

\bibitem[\protect\astroncite{{Draine}, {Roberge} \& {Dalgarno}}{1983}]{Draine:83}
{Draine}, B.~T., {Roberge}, W.~G., \& {Dalgarno}, A., 1983,
 { \apj}, { 264}, 485

\bibitem[Engargiola et al.(2003)]{Engargiola:03} 
Engargiola, G., Plambeck, R.~L., Rosolowsky, E., \& Blitz, L.,\ 2003, 
{ \apjs},{ 149}, 343 

\bibitem[Frank, King, \& Raine(2002)]{Frank:02} Frank, J., King,
A., \& Raine, D.~J.\ 2002, Accretion Power in Astrophysics
(3rd ed.; Cambridge: Cambridge Univ. Press)

\bibitem[\protect\astroncite{{Gebhardt} {\rm et~al.\/}}{2001}]{Gebhardt:01}
{Gebhardt}, K. {\rm et~ al.}, 2001,
 { \aj}, { 122}, 2469

\bibitem[\protect\astroncite{{Goldreich} \& {Tremaine}}{1979}]{Goldreich:79}
{Goldreich}, P., \& {Tremaine}, S., 1979,
 { \apj}, { 233}, 857

\bibitem[\protect\astroncite{{Gordon} {\rm et~al.\/}}{1999}]{Gordon:99}
{Gordon}, K.~D., {Hanson}, M.~M., {Clayton}, G.~C., {Rieke}, G.~H., \&
  {Misselt}, K.~A., 1999,
 { \apj}, { 519}, 165

\bibitem[Hawley, Gammie, \& Balbus(1995)]{Hawley:95} Hawley,
J.~F., Gammie, C.~F., \& Balbus, S.~A.,\ 1995, \apj, 440, 742

\bibitem[\protect\astroncite{{Kent}}{1987}]{Kent:87}
{Kent}, S.~M., 1987, 
 { \aj}, {94}, 306

\bibitem[Kim, Ostriker, \& Stone(2003)]{Kim:03} Kim, W.,
Ostriker, E.~C., \& Stone, J.~M., 2003, \apj, 599, 1157 

\bibitem[\protect\astroncite{{Kormendy}}{1993}]{Kormendy:93a} 
Kormendy, J.\ 1993, in IAU
Symp.~153, Galactic Bulges, ed. H.~Dejonghe \& H.~J.~Habing 
(Dordrecht: Kluwer) 209

\bibitem[\protect\astroncite{{Kormendy} \& {McClure}}{1993}]{Kormendy:93b}
{Kormendy}, J., \& {McClure}, R.~D., 1993,
 { \aj}, { 105}, 1793

\bibitem[\protect\astroncite{{Kulkarni} \& {Heiles}}{1987}]{Kulkarni:87}
{Kulkarni}, S.~R., \& {Heiles}, C., 1987,
 in Interstellar Processes, ed. D.~Hollenbach \& H.~Thronson (Dordrecht: Reidel), 87

\bibitem[\protect\astroncite{{Lauer} {\rm et~al.\/}}{1998}]{Lauer:98}
{Lauer}, T.~R., {Faber}, S.~M., {Ajhar}, E.~A., {Grillmair}, C.~J., \&
  {Scowen}, P.~A., 1998,
 { \aj}, { 116}, 2263

\bibitem[\protect\astroncite{{Lin} \& {Pringle}}{1987}]{Lin:87b}
{Lin}, D.~N.~C., \& {Pringle}, J.~E., 1987,
 { \mnras}, { 225}, 607

\bibitem[\protect\astroncite{{Long}, {Charles} \& {Dubus}}{2002}]{Long:02}
{Long}, K.~S., {Charles}, P.~A., \& {Dubus}, G., 2002,
 { \apj}, { 569}, 204

\bibitem[\protect\astroncite{{Mac Low} \& {McCray}}{1988}]{MacLow:88} 
{Mac Low}, M., \& McCray, R., 1988, 
 { \apj}, { 324}, 776

\bibitem[\protect\astroncite{{Martin} \& {Kennicutt}}{2001}]{Martin:01}
{Martin}, C.~L., \& {Kennicutt}, R.~C., 2001,
 { \apj}, { 555}, 301

\bibitem[\protect\astroncite{{Matthews}}{2000}]{Matthews:00} 
Matthews, L.~D., 2000, 
 { \aj}, 120, 1764

\bibitem[\protect\astroncite{{Matthews} \& {Gallagher}}{2002}]{Matthews:02}
{Matthews}, L.~D., \& {Gallagher}, J.~S., 2002,
 { \apjs}, { 141}, 429

\bibitem[\protect\astroncite{{Matthews} {\rm et~al.\/}}{1999}]{Matthews:99}
{Matthews}, L.~D. {\rm et al.}, 1999,
 { \aj}, { 118}, 208

\bibitem[\protect\astroncite{{Merritt}, {Ferrarese} \& {Joseph}}{2001}]{Merritt:01}
{Merritt}, D., {Ferrarese}, L., \& {Joseph}, C.~L., 2001,
 { Science}, { 293}, 1116

\bibitem[\protect\astroncite{{Navarro}, {Frenk} \& {White}}{1997}]{Navarro:97}
{Navarro}, J.~F., {Frenk}, C.~S., \& {White}, S.~D.~M., 1997,
 { \apj}, { 490}, 493

\bibitem[\protect\astroncite{{Quinn} \& {Goodman}}{1986}]{Quinn:86}
{Quinn}, P.~J., \& {Goodman}, J., 1986,
 { \apj}, { 309}, 472
 
\bibitem[\protect\astroncite{{Schinnerer}, {B\"oker} \& {Meier}}{2003}]{Schinnerer:03}
{Schinnerer}, E., {B{\" o}ker}, T., \& {Meier}, D.~S., 2003,
 { \apjl}, { 591}, L115

\bibitem[\protect\astroncite{{Sellwood} \& {Balbus}}{1999}]{Sellwood:99}
{Sellwood}, J.~A., \& {Balbus}, S.~A., 1999,
 { \apj}, { 511}, 660

\bibitem[\protect\astroncite{{Shakura} \& {Sunyaev}}{1973}]{Shakura:73}
{Shakura}, N.~I., \& {Sunyaev}, R.~A., 1973,
 { \aap}, { 24}, 337

\bibitem[\protect\astroncite{{Shlosman}, {Frank} \& {Begelman}}{1989}]{Shlosman:89}
{Shlosman}, I., {Frank}, J., \& {Begelman}, M.~C., 1989,
 { \nat}, { 338}, 45

\bibitem[\protect\astroncite{{Silk} \& {Norman}}{1981}]{Silk:81}
{Silk}, J. \& {Norman}, C., 1981,
 { \apj}, { 247}, 59

\bibitem[Simon et al.(2003)]{Simon:03} Simon,
J.~D., Bolatto, A.~D., Leroy, A., \& Blitz, L.\ 2003, \apj, 596, 957

\bibitem[\protect\astroncite{{Stephens} \& {Frogel}}{2002}]{Stephens:02}
{Stephens}, A.~W., \& {Frogel}, J.~A., 2002,
 { \aj}, { 124}, 2023

\bibitem[Stone et~al.(1996)]{Stone:96}
Stone, J.~M., Hawley, J.~F., Gammie, C.~F., \& Balbus, S.~A.\ 1996, \apj,
463, 656

\bibitem[\protect\astroncite{{Swaters} {\rm et~al.\/}}{2003}]{Swaters:03}
{Swaters}, R.~A., {Madore}, B.~F., {van den Bosch}, F.~C., \& {Balcells}, M.,
  2003,
 { \apj}, { 583}, 732

\bibitem[\protect\astroncite{{Toomre}}{1964}]{Toomre:64}
{Toomre}, A., 1964,
 { \apj}, { 139}, 1217

\bibitem[\protect\astroncite{{Vel\'azquez} \& {White}}{1999}]{Velazquez:99}
{Vel\'azquez}, H. \& {White}, S.~D.~M., 1999,
 { \mnras}, { 304}, 254

\bibitem[\protect\astroncite{{Wahde}, {Donner} \& {Sundelius}}{1996}]{Wahde:96}
{Wahde}, M., {Donner}, K.~J., \& {Sundelius}, B., 1996,
 { \mnras}, { 281}, 1165

\bibitem[\protect\astroncite{{Walcher} {\rm et~al.\/}}{2004}]{Walcher:04}
Walcher, C.~J., H\"aring, N., B\"oker, T., Rix, H.-W., van der Marel, R.~P., Gerssen, J., Ho, L.~C., Shields, J.~C., 2004, in Coevolution of Black Holes and Galaxies, ed. L.~C.~Ho (Pasadena: Carnegie Obs.)

\bibitem[\protect\astroncite{{Wilson} \& {Scoville}}{1989}]{Wilson:89}
{Wilson}, C.~D., \& {Scoville}, N., 1989,
 { \apj}, { 347}, 743

\bibitem[\protect\astroncite{{Wolfire} {\rm et~al.\/}}{2003}]{Wolfire:03}
{Wolfire}, M.~G., {McKee}, C.~F., {Hollenbach}, D., \& {Tielens}, A.~G.~G.~M.,
  2003,
 { \apj}, { 587}, 278

\end{thebibliography}
\end{document}